\documentclass[preprint]{aastex}
\usepackage{graphicx}

\usepackage{natbib}
\usepackage{wrapfig}

\usepackage{graphics}

\usepackage{amsmath}
\usepackage{amssymb}

\usepackage{natbib}

\newcommand{\be}{\begin{equation}}
\newcommand{\ee}{\end{equation}}
\newcommand{\nn}{\mbox{} \nonumber \\ \mbox{} }
\newcommand{\ba}{\begin{eqnarray}}
\newcommand{\ea}{\end{eqnarray}}
\newcommand{\om}{\omega}

\newcommand\eg{\textit{e.g.,\ }}

\newcommand{\Bf}{{magnetic field}}
\newcommand{\Bfs}{{magnetic fields}}

\newcommand{\NS}{neutron star}

\begin{document}

\title{Crab GeV flares from corrugated termination shock}

\author{Maxim Lyutikov\\
Department of Physics, Purdue University, \\
 525 Northwestern Avenue,
West Lafayette, IN
47907-2036 }

\author{Dinshaw Balsara \\
Department of Physics, University of Notre Dame\\
225 Nieuwland Science Hall,
 Notre Dame, IN 
46556 }

\author{Chris Matthews \\
Department of Physics, University of Notre Dame\\
225 Nieuwland Science Hall,
 Notre Dame, IN 
46556 }

\begin{abstract}
Very high energy gamma-ray  flares from the Crab nebular detected by AGILE and Fermi satellites challenge our understanding of the pulsar wind nebulae. The short duration of the flares, only few days, is particularly puzzling since it is much shorter than the dynamical times scale of the nebular. 

In this work we investigate analytically and  via numerical simulations the electromagnetic signatures expected from the large  amplitude low frequency magnetosonic waves  generated within the Crab nebular  which induce the  corrugation perturbations of the termination shock. As a result,  the oblique  termination shock produces  time-dependent,  mildly relativistic post-shock flow.  Using the relativistic MHD version of the RIEMANN code, we simulate the interaction of the termination shock with downstream perturbations.
 We demonstrate that mild Doppler boosting of the synchrotron emission  in the post-shock flow can produce  bright,  short time scale flares.
 \end{abstract}

\maketitle

\section{ The Crab gamma-ray flare: constraints from observations}
The detection of  gamma-ray flares from the Crab nebular by AGILE and Fermi satellites is one of the most astounding  recent discoveries  in high energy astrophysics \cite{2011Sci...331..736T,2011Sci...331..739A}. 
The flares typically continue for several  days/weeks. No changes in the  pulsed emission of the Crab pulsar or a  glitch 
were detected.  No variability at has been reported by other satellites:  INTEGRAL reported no detection of the flare 
the 20-400keV window (ATel\#: 2856) and Swift/BAT did not see any significant 
variability during the gamma-ray flare in the  14-150keV range (ATel\#: 2893). 
Swift also reported no evidence for active AGN near the Crab, suggesting that  the Crab itself is responsible for the flare (ATel\#: 2868). 
The prevailing conclusion from the observations of flares is that the
flares are associated with the nebular (and  not the \NS)   and are mostly likely due to the highest energy synchrotron emitting electrons. Thus,  the flares reflect the instantaneous injection/emission properties of  the nebular and are not expected to produce a noticeable change in the IC component above $\sim 1 $ GeV \citep{2010arXiv1011.4176B}.

The two most surprising properties of the flares are  intermittency  and  short time-scale variability.  For large intervals of time the gamma-ray emission form the Crab is nearly  constant, with large swings in emissivity taking place only sporadically.  This suggests  a rare,  hard to predict event that has disproportionately high-impact, a ``Black Swan" event, in the nebular as being the cause of the flare.  
The flare duration, only a few days,  is two orders of magnitude smaller than the dynamical time-scale of the nebular and several times smaller than the light crossing time of the termination shock.  This presents major problem in the interpretation of the flares as variations in the structure of the termination shock.

In present paper we consider the effects of the downstream turbulence on the properties of the termination shock.
 The nebular behaves as a resonant cavity capable of sustaining oscillations over a range of frequencies and wavelengths.  Non-linear interactions of these waves could give rise to turbulence.  A three or four $\sigma$ fluctuation in the turbulence could give rise to a strong wave that interacts with the termination shock, giving rise to an intermittent flare.  As we demonstrate analytically and via numerical simulations, for particular combination of the viewing angle, overall shock obliquity and the amplitude of the perturbing wave, the intensity of synchrotron emission in the shocked plasma can indeed experience very short spikes, on time scales much smaller than the period.

\cite{2011MNRAS.414.2017K} suggested that flares come from the so called inner knot, a Doppler-boosted emission from the 
high-velocity flow downstream 
of the oblique termination shock\footnote{We use the terms normal and oblique to indicate the relative direction of the shock normal and the fluid velocity, and the term  perpendicular for the  relative direction of the shock normal and the \Bf.}
  of the pulsar wind \citep{KomissarovLyubarsky}.  The higher resolution simulations of the Crab Nebula discovered 
strong variability of the termination shock, involving dramatic 
changes of the shock shape and inclination \citep{2009arXiv0907.3647C}.  Thus,
both observations of the morphological features of the Crab nebular \citep{2008ARA&A..46..127H}, as well as numerical simulations \citep{2006A&A...454..393B,2009arXiv0907.3647C} show dynamical behavior of times scales of months to years.
 This discovery 
suggests that the gamma-ray variability may be related to the changes in 
the characteristics of the Doppler beaming associated with this structural 
variability of the termination shock.       In this paper we study the interaction of a fast magneto-sonic wave with a relativistic MHD shock.

 Another possible scenario has been suggested by \cite{begelman_11} who posit a very rapid reconnection event.  We, however, point out that for reconnection to be rapid, it must take place within a turbulent environment \citep{lazarian_vishniac_09}.  Without picking a specific mechanism, we point out that any intermittent event with a bursty energetic release will trigger strong magneto-sonic waves that interact with the termination shock.  Owing to the indeterminacy, that interaction will be brief.

\section{Model set-up}  

In this paper we consider the  interaction of strong magnetosonic waves  generated within the bulk of the nebular with the termination shock. In Appendix  \ref{appendix} we consider a general problem of relativistic oblique  perpendicular magnetosonic shocks. The results of these calculations are used in \S \ref{out} to give simple estimates of the post-shock flow velocity and the conditions required to produce narrow high brightness peaks. 

\begin{figure}[h!]
   \includegraphics[width=0.99\linewidth]{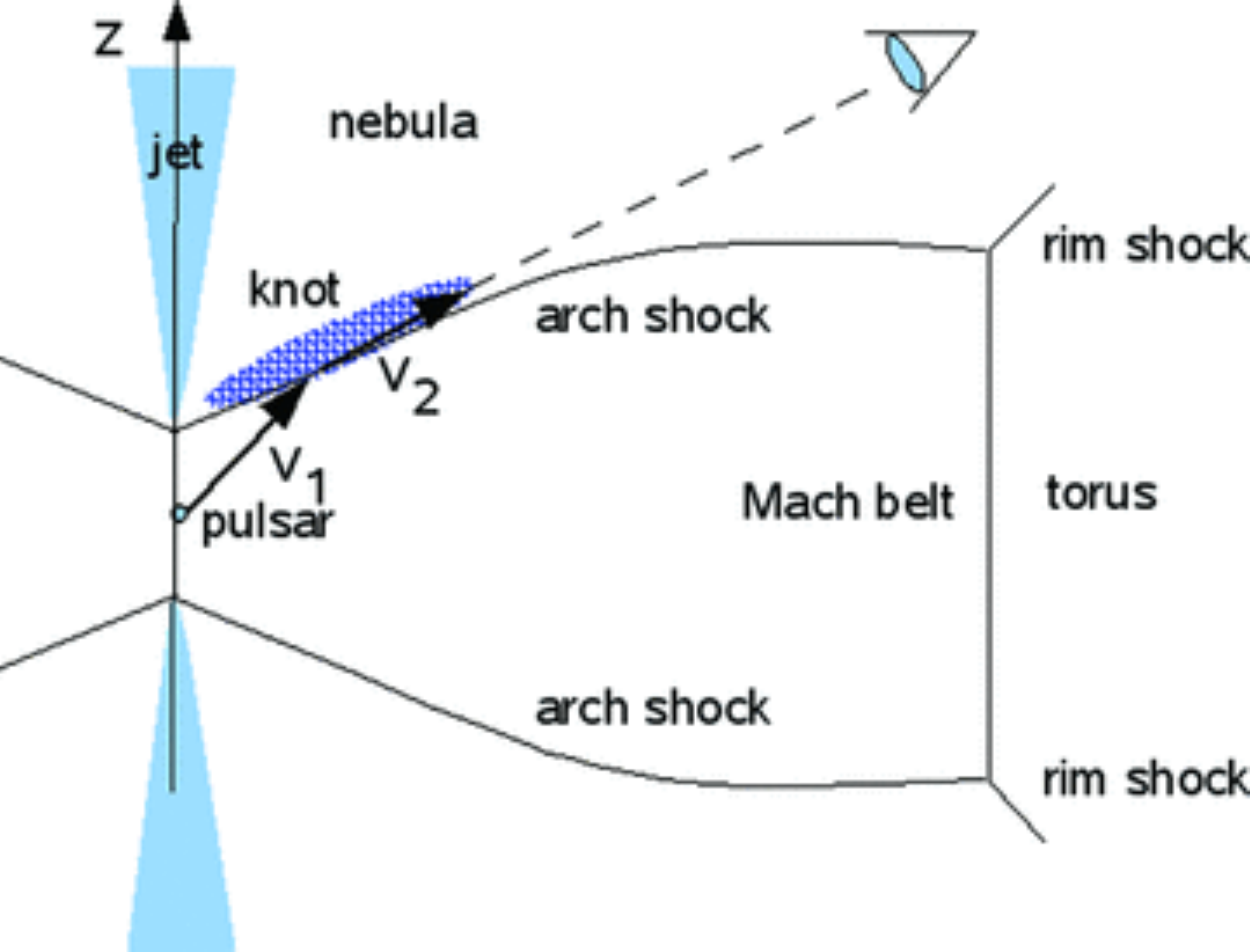}
\caption{Interpretation of the inner knot as a highly oblique part of the termination shock \citep{2011MNRAS.414.2017K}.}
\label{knot}
\end{figure}

In \S \ref{Simulations}  we describe relativistic MHD simulations of the interaction of strong magnetosonic waves  generated within the bulk of the nebular with the termination shock. Unlike  the large scale simulations of the nebular by  \citep{KomissarovLyubarsky,2009arXiv0907.3647C} we zoom-in onto the small scale details of the variability of the post-shock flow. As the underlying  steady-state flow we chose the highly oblique part of the termination shock, corresponding to the inner knot, Fig. \ref{knot} (see also Fig. 1 of \cite{2011MNRAS.414.2017K}). 
The choice of the highly oblique part of the  initial shock is important. In these parts the unperturbed post-shock flow is already mildly relativistic, with the post-shock bulk Lorentz factor of the order of $\sim 1/(\pi/2 - \theta)$. Then even mild variations of the post-shock parameters are expected to produce large variations of the Doppler factor.

  \begin{figure}[h!]
    \vskip -1.05 truein
   \includegraphics[width=0.99\linewidth]{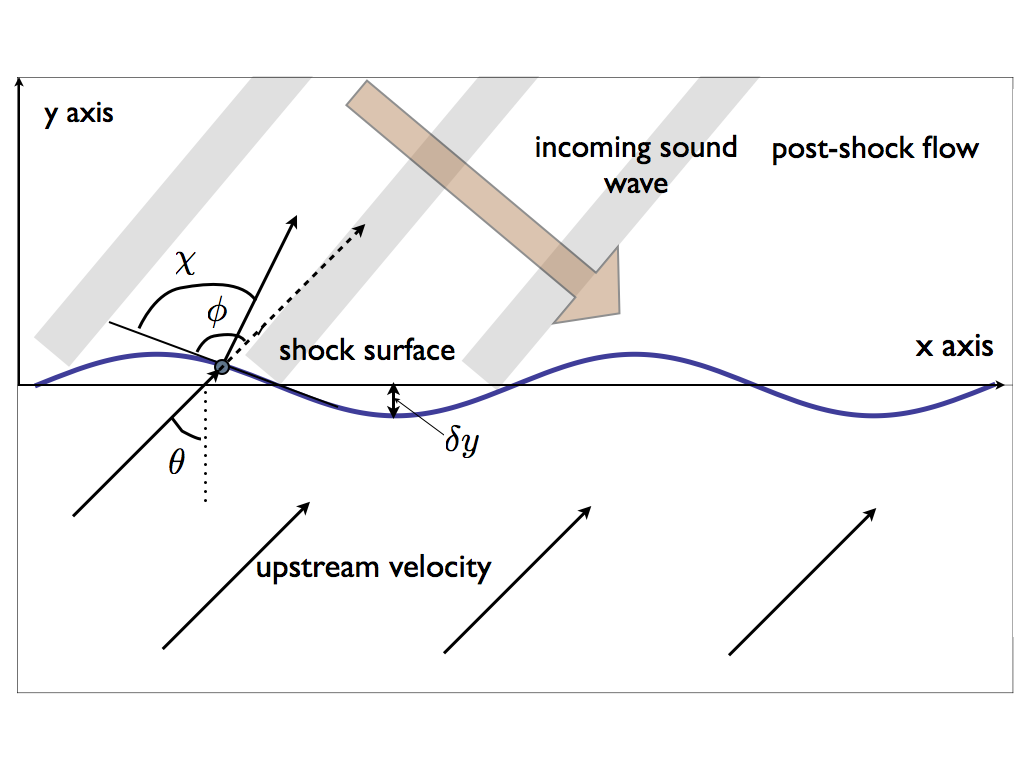}
   \caption{ Geometry of the model. Initially the  termination shock is an  oblique planar shock, so that the incoming velocity makes angle $\theta$ with the shock normal. 
Magnetosonic waves coming from the post-shock flow   induce corrugations of the shock surface (solid line). Locally,  the shock is oblique, making and angle $\phi$ with respect to the upstream velocity. The post-shock flow makes an angle $\chi$ with respect to the initial velocity (the post-shock flow is not orthogonal to the shock plane).  }
 \label{corrugation-pict-knot}
 \end{figure}

\subsection{Theoretical outline}
\label{out}

The PWN can be considered a resonator with internal frequencies of the order of the fast mode travel  time  trough the nebular. 
The resonator will support normal modes with the wave length $\sim 1/L$, the size of the nebular and frequencies $\nu \sim c/L \sim $ months-years. As  normal modes propagating in the shocked flow fall onto the termination shock,  they will be reflected back into the post-shock medium  (the reflection angle  is not equal the incoming angle, the reflected amplitude may be damped or amplified). During the reflection, the waves distort the shape of the termination shock, which becomes corrugated, see Fig.  \ref{corrugation-pict-knot}.

Studying the corrugation perturbations of shock is an important topic in modern  fluid dynamics \citep{LLVI,1984PhFl...27.1982F}.   Barring the spontaneous emission of waves by the shock\footnote{Under certain conditions  a shock can spontaneously emit a sound wave \citep{D'yakov, Kontorovich}. This requires non-trivial equation of state: in polytropic gas this does not  occur \citep{LLVI}.}  the shock corrugation can be calculated as a reflection of the sound  waves by the shock back into downstream medium.  What we are interested in here is the non-resonant response of a shock to a sound wave impinging from the shocked plasma side. That is, we are looking not for normal modes of shock oscillations, but for a response of the shock to arbitrary, non-resonant perturbation.

Since the reflected waves intensity depends on the relative phase of the incoming waves and the corrugation waves, the  corresponding perturbations of the downstream medium will include an entropy wave and one cannot use the barotropic equation of state; this complicates the problem considerably \cite[][\S 73]{1948sfsw.book.....C}. 
Still, the salient features of the interaction can be derived in the limit of small amplitude of corrugation, in which case the entropy wave is weak  \cite[][\S 73]{1948sfsw.book.....C}. In the isentropic limit, the shock acts as a partially reflecting surface  \citep[][\S 91]{LLVI}, with the angle of reflection not equal the incidence angle. The incident and the reflected wave then form an interference pattern in the shocked fluid. Propagation of the shock through this  interference pattern then induces weaker shock corrugations. 

As a simpler estimate we next consider the post-shock flow from a given corrugated shape of the shock, without calculating the dependence of the corrugation on the amplitude of the perturbing wave and neglecting the the entropy wave. General conditions at relativistic oblique perpendicular  MHD shocks are considered in 
 Appendix  \ref{appendix}.
 
\begin{figure}[h!]
   \includegraphics[width=0.49\linewidth]{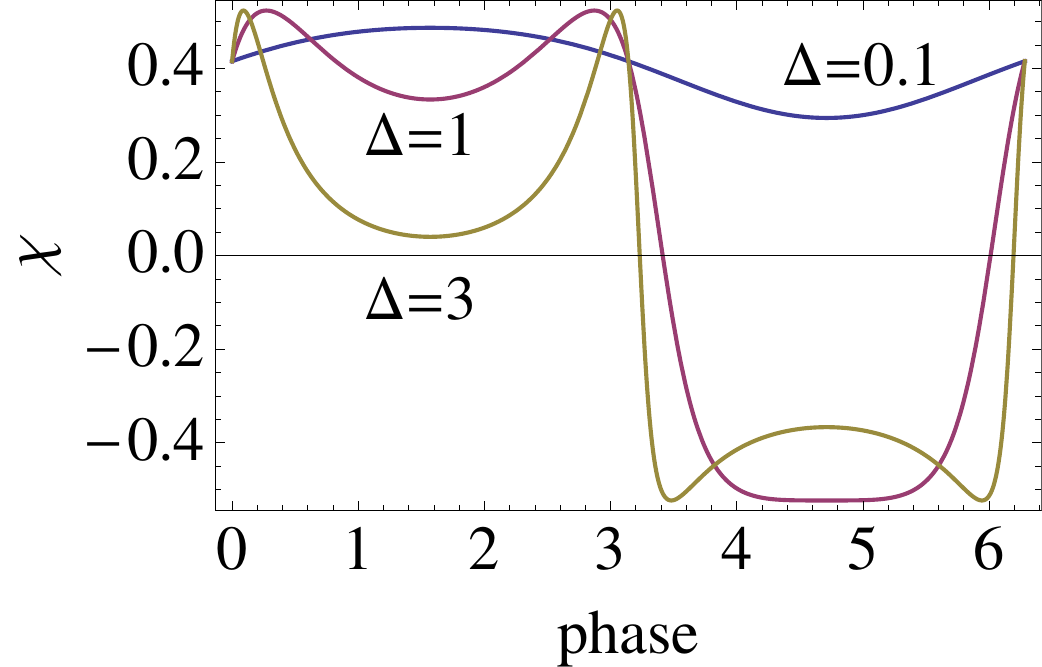}
 \includegraphics[width=0.49\linewidth]{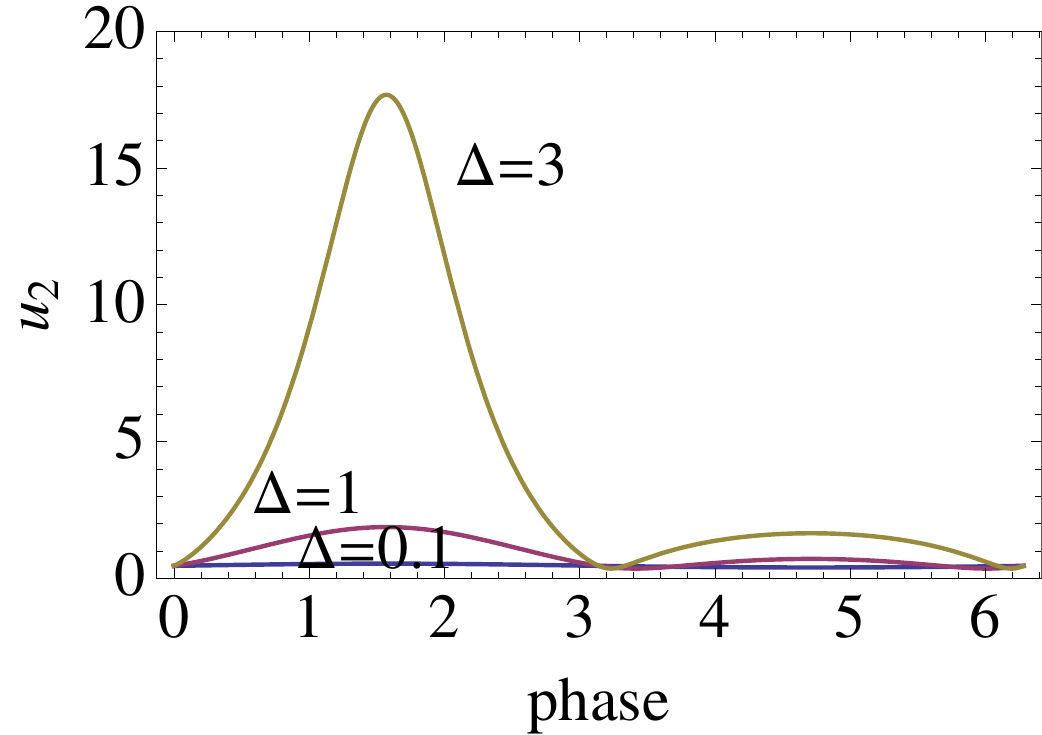}
\caption{Deflection angle $\chi$ (left panel) and post-shock momentum $u_1$ (right panel) as a function of the phase of the perturbing wave for fluid shocks $\sigma =0$ for different  wave amplitudes  $\Delta=0.1, 1, 3$. The initial obliquity angle is $\theta = 75 ^\circ$. For sufficiently high  amplitudes a given deflection angle  can be achieved up to four times per period. }
\label{post-shock}
\end{figure}

  Let us next assume that for the unperturbed oblique shock the angle of attack is $\theta$, see Fig. \ref{corrugation-pict-knot}. In addition, a wave propagating from the downstream induces shock corrugations, so that  
 in the frame of the shock  surface   is located at $y= \delta y \cos k x$, where $\delta y$ is the amplitude of the corrugation and
$k$ is the wave number of corrugation waves propagating in $x$ direction; the upstream flow is along $y$ direction. 
The angle of attack at point $x$ is then
\be
\cos \phi = { \cos \theta+ \Delta \sin k x  \sin \theta \over \sqrt{1+ (\Delta \sin k x)^2}}
\ee
where $\Delta = k \delta y$ is dimensionless amplitude of corrugations. Using the oblique shock conditions (\ref{u22}-\ref{u23}) we can then calculate the post-shock  velocity and the deflection angle, see Fig. \ref{post-shock}. The   maximum post-shock velocity  is reached at wave phase $\pi/2$ and equals 
\be
u_{2,max}= \sqrt{\frac{9 \left(\Delta ^2+1\right)}{8 (\sin \theta -\Delta  \cos \theta )^2}-1}
\ee
This value can be much larger than the initial unperturbed post-shock momentum,  $\Delta =0$,
$u_{2,0} = \sqrt{9/(8 \sin ^2 \theta) -1}$.  For mild amplitudes of corrugation, $\Delta \leq 1$, the post-shock velocity is especially high for highly oblique shocks, $\theta \approx \Delta$. 

\begin{figure}[h!]
    \includegraphics[width=0.99\linewidth]{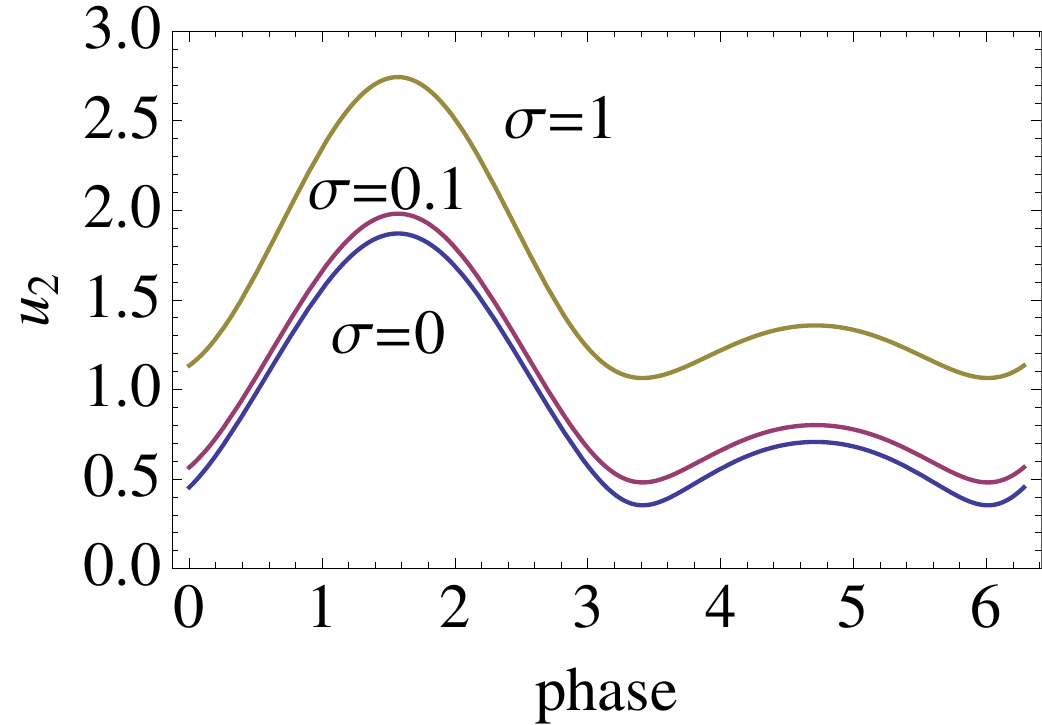}
\caption{Post-shock momentum $u_1$ as a function of the phase for different magnetization parameter $\sigma =0, 0.1, 1$ and  wave amplitudes  $\Delta=1$.}
\label{u2ofphasesigma}
\end{figure}

These estimates demonstrate that even for fluid  oblique shocks, which have substantially non-relativistic post-shock velocity for normal shocks, oblique corrugated shocks can have at specific phases very high post-shock velocities. Flows with higher magnetization have even  higher post-shock velocities, Fig. \ref{u2ofphasesigma}.

\section{Simulations}
\label{Simulations}
In \S \ref{NumMeth} we describe the numerical set up and the parameter range we investigated in our numerical simulations.  In \S \ref{numResult} we show the results of our simulations and discuss their implications.

\subsection{Numerical Methods and Simulation Parameters}
\label{NumMeth}

The simulations were run on a 480x240 zone mesh using the relativistic version of the RIEMANN code for astrophysical fluid dynamics \citep{balsara_98a}.  While the code contains higher order effects \citep{balsara_etal_09}, in this application we used a second order ADER scheme along with a HLL Riemann solver and a WENO reconstruction. The upper and lower y boundaries were periodic, the right (pre-shock) x boundary were inflow, and the left (post-shock) boundary were set to produce the inflowing magneto-sonic waves.

In the relativistic version of the RIEMANN code we evolve the conserved variables with:

\be
\partial_{t} \mathcal{U} + \partial_{i} \mathcal{F}^{i} = 0 
\ee
where
\be
 \mathcal{U}=\begin{bmatrix}\rho \\ \mathbf{S}_{j}\\U \end{bmatrix} ,   \mathcal{F}^{i} = \begin{bmatrix} \boldsymbol{v}^{i}\rho\\ \mathbf{W}^{i}_{j}\\ \mathbf{S}^{i}  \end{bmatrix}
\ee
and
\be
\mathbf {W}^{i}_{j} = \rho h \gamma^{2} \boldsymbol{v}^{i} \boldsymbol{v}_{j} - \boldsymbol{E}^{i} \boldsymbol{E}_{j} - \boldsymbol{B}^{i} \boldsymbol{B}_{j} + [P + \frac{1}{2}(E^2 +B^2)]\delta^{i}_{j}
\ee
\be
\mathbf {S} = \rho h \gamma^{2} \boldsymbol{v} + \boldsymbol{E} \times \boldsymbol{B}
\ee
\be
U = \rho h \gamma^{2} - P + \frac{1}{2}(E^2 +B^2)
\ee
We use a $\Gamma$-law equation of state which gives
\be
h = 1 + \frac{\Gamma}{\Gamma - 1}\frac{P}{\rho}
\ee
We evolve the induction equation as
\be
\partial_{t} \boldsymbol{B}^{i} + \epsilon^{i,j,k} \partial_{j} \boldsymbol{E}_{k} = 0
\ee
with ideal MHD approximation $\boldsymbol{E} = - \boldsymbol{v} \times \boldsymbol{B} $ and the magnetic field co-located at the zone boundaries in order to preserve the $\partial_{i} \boldsymbol{B}^{i} = 0$ divergence free constraint.

Here n is the Eulerian density (n=$\gamma\rho$),  $\rho$ is the Lagrangian density,  $\boldsymbol{v}$ is the three-velocity, $\mathbf{S}$ is the three-momentum, $\gamma$ is the lorentz factor, $\boldsymbol{E}$ and $\boldsymbol{B}$ are the electric and magnetic field vectors, P is the pressure, h is the enthalpy, $\Gamma $ is the polytropic index ($\Gamma = \frac{4}{3}$ in all our runs), and U is the energy density.  All vector quantities are defined in the Eulerian frame, i.e. the rest frame of the shock.  $\delta^{i}_{j}$  is the usual Kronecker delta and $\epsilon^{i,j,k}$ is the three dimensional Levi-Civita symbol. 

  This application studies the long-term interaction of waves with a stationary shock. We, therefore, require shocks to stay perfectly stationary on a computational mesh for long intervals of time when they are unperturbed. The theory of stationary relativistic MHD shocks is only reasonably useful because analytically exact solutions for such shocks are only obtained in the limit of vanishing pre-shock pressure. Using the exact solutions calculated using the jump conditions in Appendix  \ref{appendix} as an approximate starting point, we evolved the solution with finite but small pre-shock pressures on a very long one dimensional computational mesh. Finite pressure and discretization errors do make this shock move slowly with respect to the mesh. By running these one dimensional simulations for a very long time, we were able to quantify the speed with which the shock drifts on the long mesh. We then lorenz boosted the fluid variables into the rest frame of the shock. The resultant densities, pressures, velocities and magnetic fields in the pre- and post-shock regions for the stationary shocks have been tabulated in Table \ref{Tab1}. 

 The resulting numerically stable post-shock values were used to calculate the eigenvector for the fast right-going magneto-sonic wave using the analytical procedure from\citep{balsara_01}. The resulting wave was scaled so the density perturbation was a fraction of the unperturbed post-shock density, and then allowed to propagate into the shock at a 45$^{o}$ angle to the shock normal. We explored a range of angles of attack ($\phi$) to test the effect of the obliqueness of the shock, and magnetization ($\sigma$) to examine the effect of the magnetic field strength in the pulsar wind. We also varied the amplitude ( $\delta\rho/{\rho_{1}}$) of the perturbing magneto-sonic waves.

Table \ref{Tab1} describes the primitive variables we used to set up the shocks. In all cases the frame has been rotated so that the shock is in the y-z plane, with the magnetic field in the positive z direction. The odd numbered runs had perturbation strengths of ${\delta\rho}/{\rho_{1}}=0.3$ while the even numbered runs used ${\delta\rho}/{\rho_{1}}=0.7$. Table \ref{Tab2} gives the parameters defined in Appendix  \ref{appendix} corresponding to these choices of primitive variables and compares the deflection angles and compression ratios calculated using the numerical code and the analytical expressions.

\begin{table}[h!]
\caption{\label{Tab1}Run parameters used to set up the un-perturbed shocks for our simulations.  $\rho$ is the density in the Lagrangian frame of the fluid, v$_{x}$ is the fluid velocity normal to the shock, v$_{y}$ is the fluid velocity along the shock, and B$_{z}$ is the magnetic field perpendicular to the shock normal and the velocity vector.  The velocities and magnetic field are all given in the rest frame of the shock.}
\begin{tabular*}{.95\textwidth}{@{\extracolsep{\fill}}|c|c|c|c|c|c|c|c|c|c|}
\hline
Run&$\rho_{0}$& v$_{x0}$&v$_{y}$&P$_{0}$&B$_{z0}$&$\rho_{1}$& v$_{x1}$&P$_{1}$&B$_{z1}$ \\ \hline
1,2& 0.1118& -0.8452& 0.5286& 0.001& 0.2878& 3.8050& -0.2542& 9.100& 0.951 \\ \hline
3,4& 0.1047& -0.7007& 0.7057& 0.001& 0.9363& 1.720& -0.2674& 1.948& 2.466 \\ \hline
5,6& 0.1190& -0.3482& 0.9270& 0.001& 1.1254& 0.6680& -0.1538& 0.152& 2.5610\\ \hline
7,8& 0.1144& -0.9831& 0.1698& 0.001& 0.2946& 5.011& -0.3054& 16.1850& 0.9510 \\ \hline
9,10& 0.1059& -0.9523& 0.283& 0.001& 0.9468& 1.9960& -0.3860& 2.8140& 2.3380 \\ \hline
11,12& 0.1488& -0.8937& 0.4150& 0.001& 1.0683& 1.1446& -0.4373& 0.724& 2.1810 \\ \hline
\end{tabular*}
\end{table}

\begin{table}[h!]
\caption{\label{Tab2}Variables describing the geometry of the un-perturbed shocks.  $\phi$ is the angle of attack, $\chi$ is the deflection angle, $\sigma$ is the magnetization of the pre-shock plasma, $\eta$ is the compression ratio between the post- and pre-shock fluids, and $\gamma_{0,1}$ are the Lorentz factors for the pre- and post-shock plasmas respectively.  The table also compares  the deflection angles $\chi_{num}$ and compression ratios  $\eta_{num}$calculated using the numerical code and the analytical expressions, $\chi_{an}$ and $\eta_{an}$, calculated using Appendix  \ref{appendix}.  }
\begin{tabular*}{.95\textwidth}{@{\extracolsep{\fill}}|c|c|c|c|c|c|c|c|c|c|}
\hline
Run&$\phi$& $\chi_{num}$&$\chi_{an}$&$\sigma$&$\eta_{num}$& $\eta_{an}$&$\gamma_{0}$&$\gamma_{1}$\\ \hline
1,2& 57.9776$^{o}$& 32.2950$^{o}$& 32.1166$^{o}$& 0.0046& 32.2458& 33.850 & 12.6806& 1.2347\\ \hline
3,4& 44.7963$^{o}$& 24.0439$^{o}$& 24.0094$^{o}$& 0.0922& 16.4279&16.359& 9.5316& 1.5241 \\ \hline
5,6& 20.5872$^{o}$& 11.1670$^{o}$& 11.052$^{o}$& 0.2068& 5.6134& 5.474& 7.1745& 2.9233\\ \hline
7,8& 80.2006$^{o}$& 19.2743$^{o}$& 19.324$^{o}$& 0.0036& 43.8024& 44.1785& 14.6140& 1.0673\\ \hline
9,10& 73.4494$^{o}$& 19.6968$^{o}$& 19.7816$^{o}$& 0.1103& 18.8480& 19.042 & 8.7586& 1.1390\\ \hline
11,12& 65.0917$^{o}$& 18.5929$^{o}$& 19.9364$^{o}$& 0.2230& 7.6922&10.1557 & 5.8646& 1.2534 \\ \hline
\end{tabular*}
\end{table}

\subsection{Simulation Results and Analysis}
\label{numResult}
 \begin{figure}[h!]
    \includegraphics[width=0.70\linewidth]{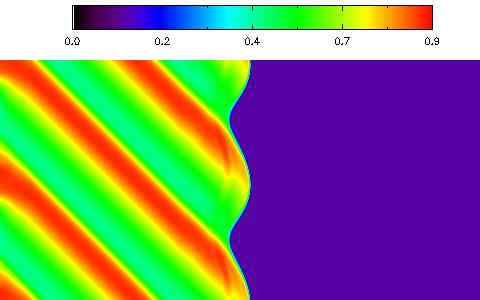}
\caption{Density for run 5 at early time, shortly after the wave hits the shock.  The corrugation wave in the shock front is especially pronounced at these early times. }
\label{Run_5_rho}
\end{figure}

 \begin{figure}[h!]
    \includegraphics[width=0.70\linewidth]{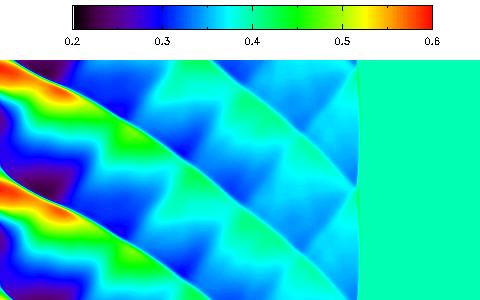}
\caption{y-velocity (transverse velocity) for Run 12 at later time.  The reflected waves are interacting with the incoming waves and creating this fish-scale pattern.}
\label{Run_12_vly}
\end{figure}

Our simulations all show initially strong corrugations when the fast magneto-sonic wave first impinges on the shock.   Figure \ref{Run_5_rho} shows the density for our Run 5 shortly after the incoming wave has hit the shock, clearly showing the corrugation of the shock front.  Larger perturbation amplitude produces more significant and persistent corrugation of the shocks.  In extreme cases the perturbations could even drive the shock front off of the computational domain.  These perturbations effectively constitute an extra ram-pressure term, which increases the effective post-shock pressure, moving the shock to the right.  In most situations, however, the perturbations are gentle and the shocks either show a slow, secular rightward motion or none at all.

   As time progresses the reflected waves coming off the shock interfere with the incoming waves, and the corrugation of the shock front begins to damp away.  This damping is due to the homogenization of the effective ram pressure term as many wave fronts hit the shock.  Once the simulation has stabilized we see a prominent fish-scale pattern formed by the incoming and reflected waves, with the largest variations in the fluid variables occurring at the intersections of the waves.  Figure \ref{Run_12_vly} provides an example of this, showing the transverse velocity in Run 12, after the reflected waves have had time to fully cross the post-shock fluid.

In all our runs we see the strongest fluctuations in the flow variables, and correspondingly the strongest variation in the observed intensity, at the early times when the shock is most deformed by the impinging magneto-sonic waves.  After the shock front has had time to regain equilibrium the variation in the Lorentz factor is smaller, and we do not see the brief spikes in intensity from Doppler boosting.

\section{Modeling the Gamma-ray Emission}
\label{HE_signal}

In  this section we discuss the factors responsible for the observed high energy gamma-rays in the instantaneous approximation. Later,   in \S \ref{ToF_effects} we discuss the effect of time-of-flight considerations on the observed intensity.

The highest energy leptons emitting synchrotron radiation at $\sim 100 $ MeV have a  very short cooling time scale, of the order of $t_{\rm cool} \sim 10^5 \, \gamma^3\, {\rm sec} $, where $\gamma$ is the bulk factor of the post-shock flow and the estimate is done for the inner wisp located at $\sim 10^{16}$ cm.  Thus, we expect that only a fairly narrow region behind the shock contributes to the very high energy emission.
In addition, the observed radiation is highly modulated by the relativistic beaming effects. One expects that intrinsic variations of proper emissivity have only a marginal influence on the observed flux: it is mostly dominated by Doppler boosting.

As a simple prescription that takes into account the radiative decay of shock particles and the Doppler boosting we integrate  along the line of sight the quantity $\delta ^4$ (here $\delta = 1/( \gamma (1-\beta \cos\theta_i))$ is the Doppler factor, $\theta_i$ is the instantaneous angle between line of sight and the fluids velocity) for a fixed downstream  distance from the shock. This is a simplified prescription that captures the two salient features of the post-shock flow: fast radiative decays and the dominant effects of relativistic Doppler boosting. 

\begin{figure}[h!]
\includegraphics[width=0.70\linewidth]{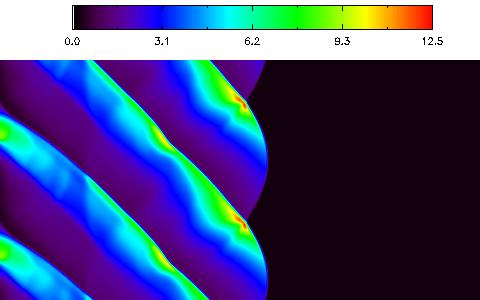}
\caption{Synchrotron emissivity (I$\propto$P*B$^{\alpha}$) for Run 12 at early time with $\alpha$=1.5, calculated in the fluids rest frame. The effect of the variation in this proper emissivity is expected to be negligible compared to the effects of the angle dependent Doppler boosting.}
\label{Run_12_sync4}
\end{figure}

Figure \ref{Run_12_sync4} shows a rough estimate for the intrinsic  synchrotron emissivity in Run 12, calculated using the approximation I$\propto$P*B$^{\alpha}$, with $\alpha$=1.5 as in \cite{jun_jones_99,balsara_etal_01}.  As with the other variables, we see the highest intensity in the regions around the intersection of the incoming and reflected waves.  While synchrotron radiation is expected to produce the high energy gamma-rays, the effect of the line of sight dependent Doppler boosting should be the dominant term determining the observed intensity.

 \begin{figure}[h!]
    \includegraphics[width=0.70\linewidth]{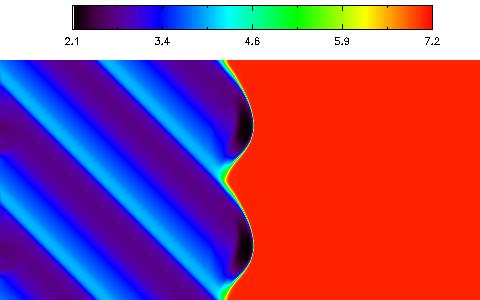}
\caption{Lorentz factor for Run 6 at early time.  The strong variation (a factor of 2) in the post-shock Lorentz factor leads to strong variation in the Doppler boosting along some lines of sight, which leads to spikes in the intensity.}
\label{Run_5_lor}
\end{figure}

\begin{figure}[h!]
    \includegraphics[width=0.70\linewidth]{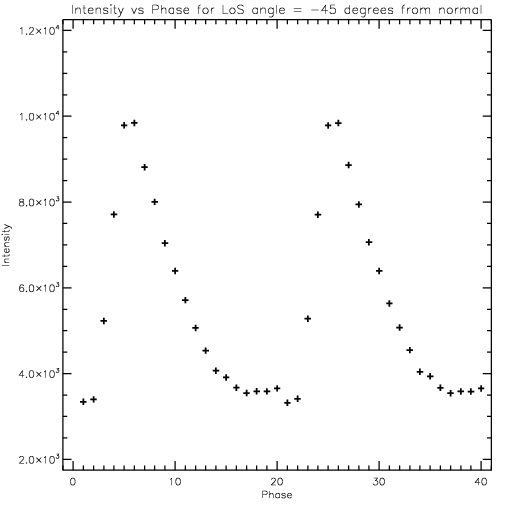}
\caption{Calculated intensity as a function of the phase of the corrugation wave for Run 12 at early time, with the line of sight angle at -45$^{o}$ to the  shock normal.  This profile assumes a constant rest frame emissivity, and only reflects the effect of the LoS dependent Doppler boosting ( I$\propto$$\delta ^4$ ) }
\label{Run_12_int}
\end{figure}

Figure \ref{Run_5_lor} shows the Lorentz factor for Run 6, which represents our most oblique parameters ($\phi = 20.59^{o}$) with strongest perturbations ($\delta\rho/\rho_{1}$= 0.7).  One can see that the perturbations in the post-shock flow (with unperturbed Lorentz factor $\gamma_{1}$= 2.9 ) are large, ranging from 2.09 to 3.96.  This large variation in the post-shock  Lorentz factor can lead to strongly enhanced observed intensity at some preferred lines of sight angles due to Doppler boosting, see Fig. \ref{Run_12_int}.

In the relative intensity profiles calculated using this simple radiation model described above, we see generally broad profiles for most line of sight angles, with a few angles producing sharp peaks from small sections of the corrugation wave (Fig. \ref{Run_12_int}).  These peaks could be as high as twice the baseline intensity.  Since the peaks are related to small segments of the shock front, they have a small characteristic time scale.  These flares are most pronounced at early times of the wave-shock interaction,  when the shock is strongly corrugated.

\section{Time-of-flight effects: Cherenkov-type interference}

\label{ToF_effects}

The main drawback of the simplified radiation modeling described above is that it does not take into account the time-of-flight effects. One expects that neglecting the time-of-flight effects smooths out the intensity variations. We expect that the full temporal analysis will produce even sharper peaks: there is a special combination of the incoming wave direction and the amplitude of the corrugation such that the emission from the particular point of the corrugation wave add up coherently. Let us demonstrate this for a simple case of initially normal fluid shock.

Consider a wave falling from downstream onto  ? relativistically strong normal shock at an angle $\alpha$ to the shock normal.  In the moving fluid sound waves become dispersive. For an unperturbed  fluid moving relativistically with velocity $v_1=1/3$ in the medium with sound speed $c_s=1/\sqrt{3}$, the corresponding dispersion relation
\citep[][Eq. 10.18]{Anile:1989} becomes
\be
{\om \over k} ={ \sqrt{ 50 - 2 \cos 2\alpha} -  3 \cos \alpha \over 13}
\ee
The phase speed of corrugation perturbations along the shock normal is then
\be
v_{ph} =  { \sqrt{ 50 - 2 \cos 2\alpha} -  3 \cos \alpha \over 13 \sin \alpha}
\label{vph}
\ee
The corrugations become luminal for incidence angles smaller than $\alpha= 0.32$. For smaller incidence angles the corrugation wave is superluminal. If a post-shock flow  at some phase satisfies the Cherenkov condition
\be
\sin \chi = c/v_{ph}
\label{chere}
\ee
the waves emitted by this phase will add up constructively. Equation (\ref{chere}) together with 
the expression for the phase speed (\ref{vph}) and  the post shock flow angle (\ref{u23}) with $\phi=\pi/2$ defines the condition on the  incidence angle of a sound wave coming from the shocked fluid, such that the resulting corrugations of the shock surface 
produce the post-shock flow which remains in phase with the corrugation wave. 
 In this case    the corrugation pattern acts as a real particle, emitting light at the Cherenkov condition.

\section{Discussion}

In this work we considered  emission from a  relativistic corrugated  shocks. We performed relativistic simulations of the corrugations of the perpendicular MHD shocks induced  by strong magnetosonic waves impinging from the downstream medium. Our main conclusions is that long wavelength perturbations with the relative amplitude   of $\sim 0.5 - 0.7$ of the downstream density can result in sharp intensity variations, with duration  of the peak much smaller than the wave period.  Sharp intensity variations can have amplitudes of  an order of magnitude in flux, with large modulation index, of the order of $0.5$. 

Variable, mildly relativistic post-shock flow may also explain (or, in fact,  be required in stochastic  shock acceleration models) the very high break energy  observed in the Crab flares. As argued by \cite{2010MNRAS.405.1809L} \citep[see also][]{1996ApJ...457..253D}, there is an acceleration model-independent  upper limit on the frequency of synchrotron emission by radiation reaction-limited  acceleration of electrons:
\be
\epsilon_{\rm max} \approx  \hbar  { m c^3 \over e^2} \approx 100 \mbox{ MeV}.
\label{emax}
\ee
The break energy observed by {\it Fermi} satellite   in Crab's  quiescent state,  $\sim 100$ MeV \citep{2009arXiv0909.0862F}, approaches this limit, while the break energies  during flares, $\sim 500$ MeV \citep{2011Sci...331..736T,2011Sci...331..739A}, exceed it. If emission is generated  in a relativistically moving plasma,  the  maximum observed synchrotron frequency is boosted by the Doppler factor. Note also, that previously EGRET data did indicate a moderate level of the cut-off energy  variability \citep{1996ApJ...457..253D}.) 

Another important feature of our model is that  in  the proposed model the time scale of the flare is set not by the radiative decay time of the emitting particles, but by the overall dynamics of the termination shock. Thus, the flare time scale cannot be used to estimate the \Bf\ within the emission region. 

We would like to thank Sergey Komissarov for suggesting  the possibility discussed in this paper.

\bibliographystyle{apj}
  \bibliography{/Users/maxim/Home/Research/BibTex}

\appendix
\section{Oblique  relativistic MHD shocks}
\label{appendix}

We  derive jump conditions at oblique relativistic perpendicular MHD shocks, when  the \Bf\ is orthogonal to flow velocity which is not aligned with the shock normal. We investigate  the post-shock parameters as functions of the upstream plasma magnetization $\sigma$, Lorentz factor $\gamma_1$  and the angle of attack $\phi$. 
Previously, \cite{1987JPlPh..37..117W} investigated oblique  relativistic MHD shocks in the quasi-parallel case, when the planes of  fluid velocity and of the \Bf\ coincide.  \cite{1980PhFl...23.1083K} considered relativistic  fluid  oblique shocks. Here we consider relativistic  oblique perpendicular  magnetohydrodynamic  shocks, so that the  \Bf\ is in the plane of the shock and is orthogonal to the fluid velocity. This appendix corrects an error in \cite{lyutikov04}.

\begin{figure}[ht]
\includegraphics[width=0.8\linewidth]{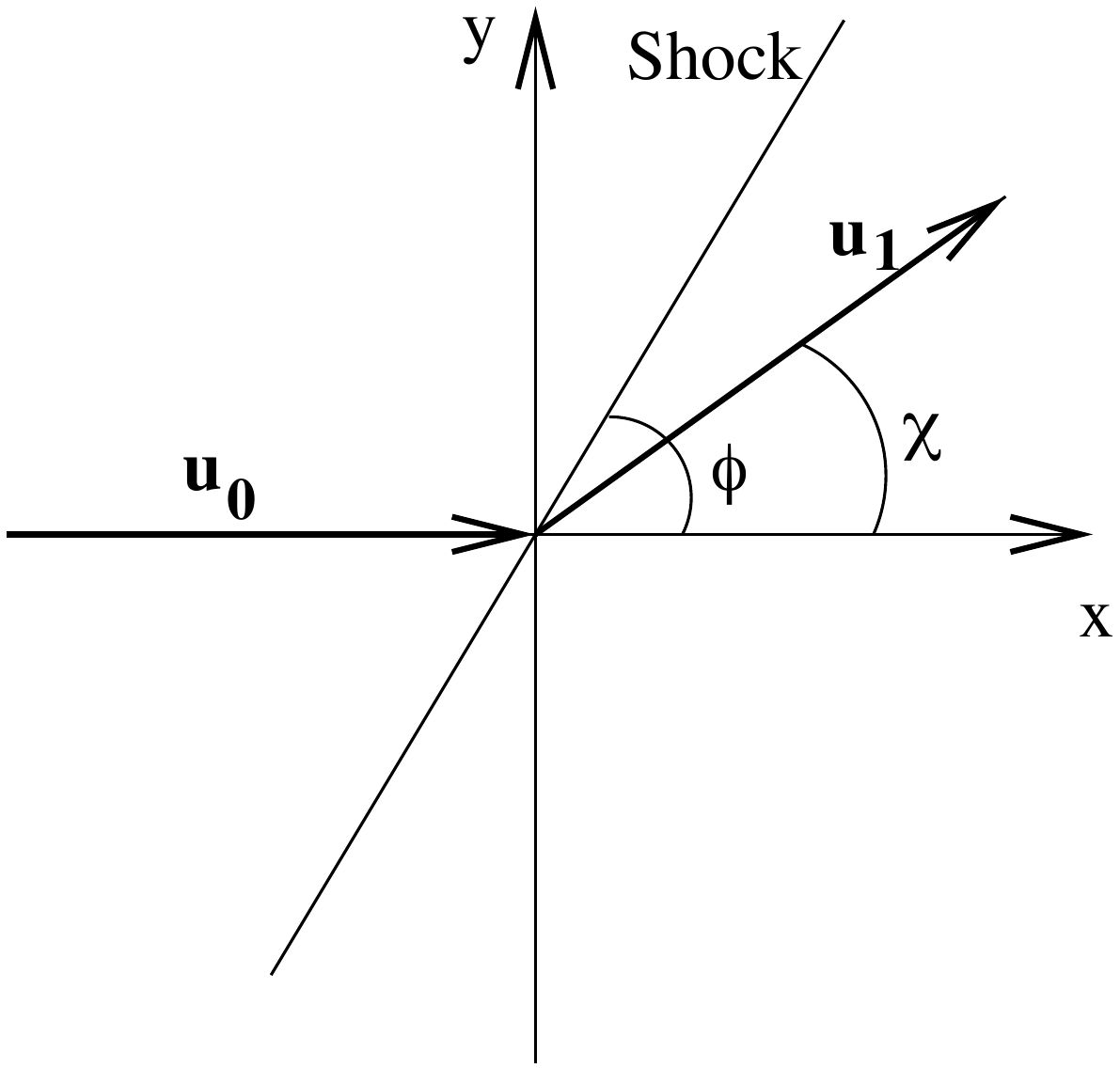}
\caption{Geometry of the flow}
\label{geom}
\end{figure}

Let the stream lines make an angle $\phi$ with the shock normal
and let the post-shock flow
make  an angle $\chi$ with the initial velocity, Fig. \ref{geom}.  Normal shock corresponds to $\phi = \pi/2$. 

Keeping only $x$-dependence, the equations of perpendicular relativistic MHD read \citep[\eg][]{Anile:1989}
\ba &&
\partial_x \left( n u_x  \right) =0
\nn &&
\partial_x  \left(p  +B^2/2+ ( w +B^2)  u_x^2  \right)=0
\nn &&
 \partial_x  \left( (w +B^2) \gamma  u_x\right)=0
 \nn &&
 \partial_x \left( u B\right) =0
 \ea
where $n$ is proper density, $w=e +p=n+\Gamma/(\Gamma-1) p$ is fluid enthalpy, $e$ is internal energy, $B$ is rest-frame \Bfs, $u$ is a four-velocity. Note that in this Appendix  we use  the  rest-frame \Bf\ and density as primitive variables, while the numerical simulations in \S \ref{NumMeth} use the quantities  in the lab frame. 

Oblique shock conditions can be obtained from normal shocks and an additional condition
that the component along the shock velocity remains constant \cite[\eg][]{1980PhFl...23.1083K,Anile:1989}:
\ba && 
n_0 u_0 \sin \phi  = n_1 u_1 \sin (\phi - \chi)
\nn &&
B_0 u_0 \sin \phi  = B_1 u_1 \sin (\phi - \chi)
\nn &&
p_0  +B_0^2/2+ ( w_0 +B_0^2)  u_0^2 \sin^2  \phi   = p_1 +B_1^2/2 + (  w_1 +B_1^2)  u_1^2 \sin^2   (\phi - \chi)
\nn &&
( w_0 +B_0^2) \gamma_0  u_0 \sin \phi  =(  w_1 +B_1^2)  \gamma_1  u_1 \sin  (\phi - \chi)
\nn &&
\beta_0 \cos \phi  = \beta_1 \cos  (\phi - \chi)
\label{1}
\ea
where  subscripts refer to upstream ($0$) and downstream ($1$) regions and  $\beta$ is three-velocity.
 Below we assume the  relativistic equations of  state with $\Gamma=4/3$.

The problem is to express the post-shock parameters ($p_1,n_1, B_1, u_1$ and $\chi$) in terms of the upstream parameters   ($p_0,n_0, B_0, u_0$ and $\phi$). As we demonstrate below, the corresponding equation take especially simple form is expressed in terms of the compression ratio 
$
\eta = n_1/n_0
$, which in turn  depends on the upstream parameters.   By ideal condition
$ 
B_1 = \eta B_0
$. Compression ratio can be related to the turning angle $\chi$
\be
\eta^2 = \left( 1+u_0^2 \sin ^2 \phi - \gamma_0^2 \sin ^2  (\phi - \chi)\right) {\tan^2 \phi \over \sin ^2  (\phi - \chi)}
\label{etaofchi}
\ee

 In term of $\eta$,
\ba &&
u_0 ^2  \sin ^2 \phi = {\left( ( p_1-p_0 ) +(B_1^2 -B_0^2)/2  \right)\eta^2\over \eta^2 (  w_0+ B_0^2) - ( w_1+ B_1^2)}
\nn &&
u_1   \sin ^2  (\phi - \chi) 
 =  {\left( ( p_1-p_0 ) +(B_1^2 -B_0^2)/2  \right)\eta^2\over \eta^2 (  w_0+ B_0^2) - ( w_1+ B_1^2)}
  \nn &&
u_1 ^2 = { u_0^2 \over \eta^2} { \sin^2  \phi  \over \sin ^2  (\phi - \chi) } = { \eta^2 \cos  ^2 \phi +\gamma_0^2 \sin  ^2 \phi \over \eta ( 1+ u_0^2 \sin  ^2 \phi}
\nn && 
\sin ^2  (\phi - \chi)= { \sin^2 \phi \over \eta^2 \cos ^2 \phi + \sin  ^2 \phi }
 \ea

Introducing magnetization parameter $\sigma = B_0^2 /w_0$, the jump conditions (\ref{1})  can be resolved for post-shock pressure  $p_1$ and enthalpy  $w_1$ as functions of $\eta$: 
\ba && 
p_1 = p_0 - { w_0 \sigma (\eta^2 -1) \over 2} + u_0 ^2 w_0 \sin ^2 \phi  (1+\sigma) \left( 1-  \sqrt{ 1+u_0^2 \sin ^2 \phi \over \eta^2 + u_0^2 \sin ^2 \phi } \right)
\nn  && 
w_1 =  w_0 \eta^2\left(  -\sigma +(1+\sigma)  \sqrt{ 1+u_0^2 \sin ^2 \phi \over \eta^2 + u_0^2 \sin ^2 \phi } \right) 
\label{x}
 \ea
 
 In case of cold plasma, $p_0=0$ the compression ratio $\eta$ can then be simply  related to the inflow parameters:
 \be
\eta ^2 \left((\sigma +1) \sqrt{\frac{u_0^2 \sin ^2 \phi +1}{u_0^2 \sin ^2 \phi +\eta ^2}}+\sigma \right)+4 u_0^2 (\sigma +1) \sin
   ^2 \phi  \left(\sqrt{\frac{u_0^2 \sin ^2 \phi +1}{u_0^2 \sin ^2 \phi +\eta ^2}}-1\right)-\eta -2 \sigma=0
    \label{etasigma}
    \ee
 
\begin{figure}[h!]
\includegraphics[width=\linewidth]{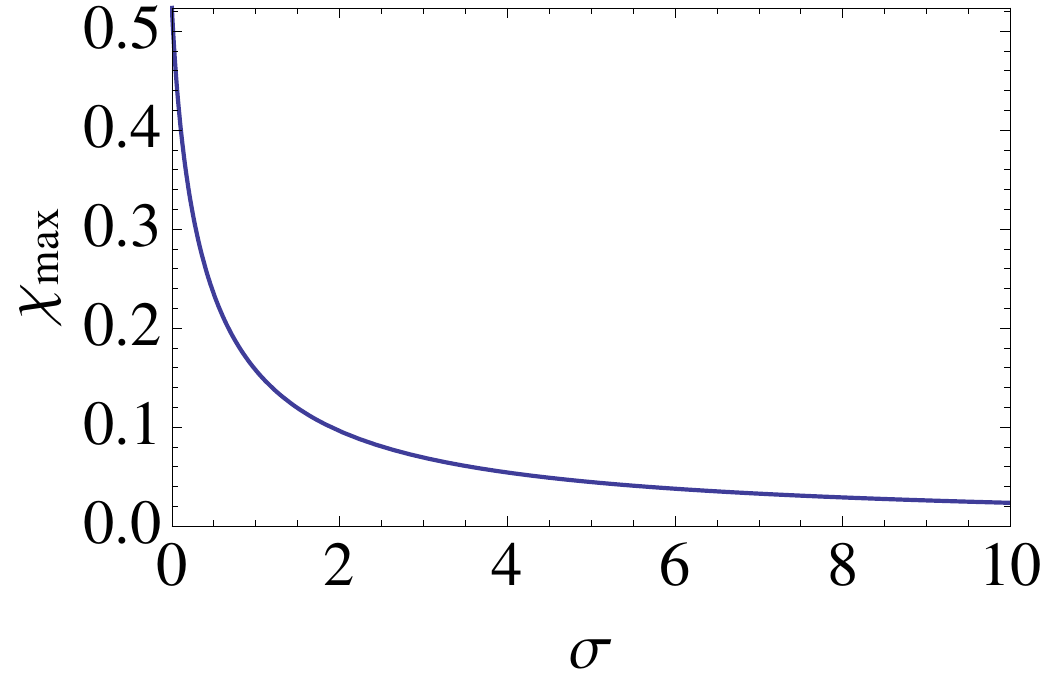}
\caption{Maximum  deflection angle $\chi_{max} $ as a function of the  magnetization parameters $\sigma$ in the limit $u_0^2 \rightarrow \infty$. $\chi_{max}=\pi/6$ for $\sigma =0$,   $\chi_{max} \approx 1/(4 \sigma) $ for $\sigma \gg 1$. }
\label{chimax-sigma}
\end{figure}

 Let us next simplify the above relations  in the high compression limit,  $\eta \rightarrow \infty$.  (The strong compression limit becomes  inapplicable for $\sin \phi  \leq 1/( u_0 \sqrt{1+\sigma})$.)  Eq. (\ref{etasigma}) gives 
 \be
 { \eta^2 \over u_0^2}  = \frac{16 (1+\sigma) \sin ^2 \phi }{8 \sigma ^2+2 \left(\sqrt{16 \sigma ^2+16 1+\sigma}+5\right) \sigma +\sqrt{16 \sigma ^2+16 \sigma
   +1}+1}  \approx
   \left\{  \begin{array}{cc}
   8 (1- 9 \sigma)  \sin ^2 \phi, & \sigma \ll 1\\
   {1   \sin ^2 \phi \over \sigma}, &  \sigma  \gg 1
   \end{array}
   \right.
   \ee
   The turning angle is given by
   \ba &&
  \tan \chi = { (1 - {\cal Z}) \cos \phi \sin \phi \over \cos^2 \phi +  {\cal Z}  \sin ^2 \phi}
  \nn &&
   {\cal Z} =\sqrt{\frac{8 \sigma ^2+2 \left(\sqrt{16 \sigma ^2+16 \sigma +1}+5\right) \sigma +\sqrt{16 \sigma ^2+16 \sigma +1}+1}{8 \sigma ^2+2
   \left(\sqrt{16 \sigma ^2+16 \sigma +1}+13\right) \sigma +\sqrt{16 \sigma ^2+16 \sigma +1}+17}}, \,  1/3 <   {\cal Z} < 1
   \ea
   Maximum deflection angle is reached at $\cos \phi = \sqrt{  {\cal Z} /(1+ {\cal Z} }$ and equals $\tan \chi_{max} = 
   (1 - {\cal Z}) /(2 {\cal Z}) =  1/2( \tan \phi -\cot \phi) $, Fig. \ref{chimax-sigma}.

\begin{figure}[h!]
\includegraphics[width=\linewidth]{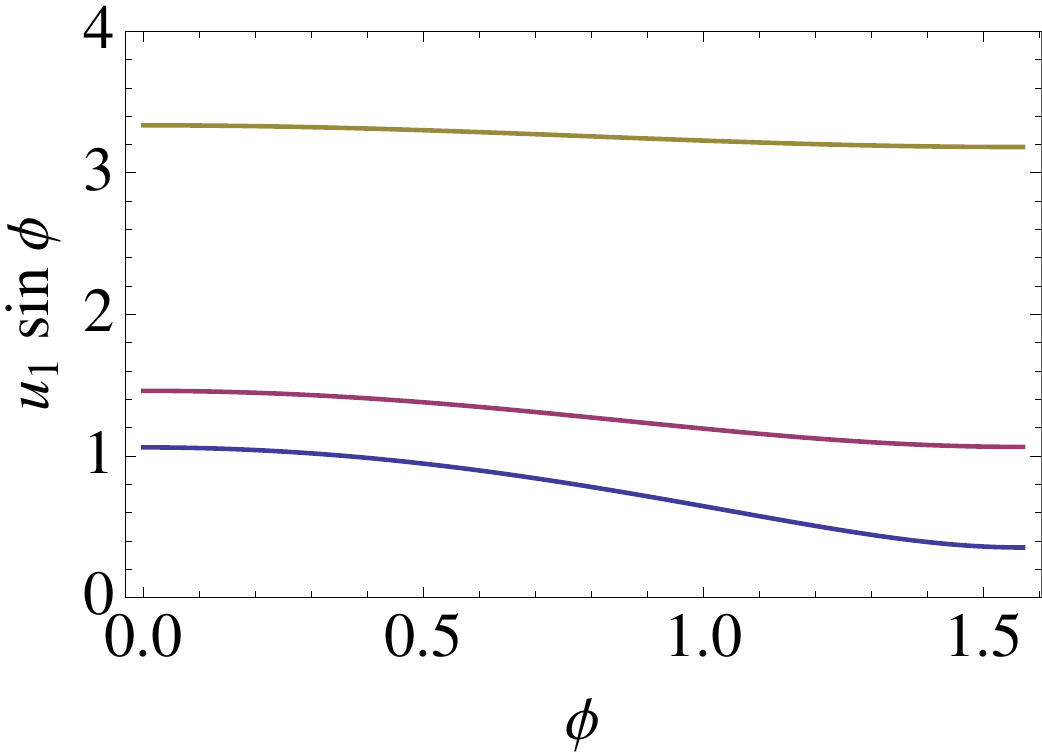}
\caption{Dependence of   $u_1 \sin \phi$  (the post-shock four-velocity times angle of attack) the angle of attack $\phi$ for different magnetization parameters $\sigma=0,1,10$ (bottom to top). }
\label{u222}
\end{figure}

    The post-shock four-velocity
    \be
    u_1^2  \sin^2 \phi =\frac{1}{8}
   \left(9-8 \sin ^2\phi \right) 
+
 \frac{8 \sigma ^2+2 \left(\sqrt{16 \sigma ^2+16 \sigma +1}+4\right) \sigma +\sqrt{16 \sigma ^2+16 \sigma +1}-1}{16 (\sigma +1)}   \label{u22}
   \ee
   \citep[for normal shock, $\phi=\pi/2$,  Eq. (\ref{u22}) reproduces Eq. (4.11) of ][]{kc84}, see Fig. \ref{u222}.
In the high $\sigma$ limit, $u_1^2   \sin^2 \phi \approx \sigma   + ( 9/8 -\sin^2 \phi) + {\cal O} (1/\sigma)$.

\begin{figure}[h!]
\includegraphics[width=\linewidth]{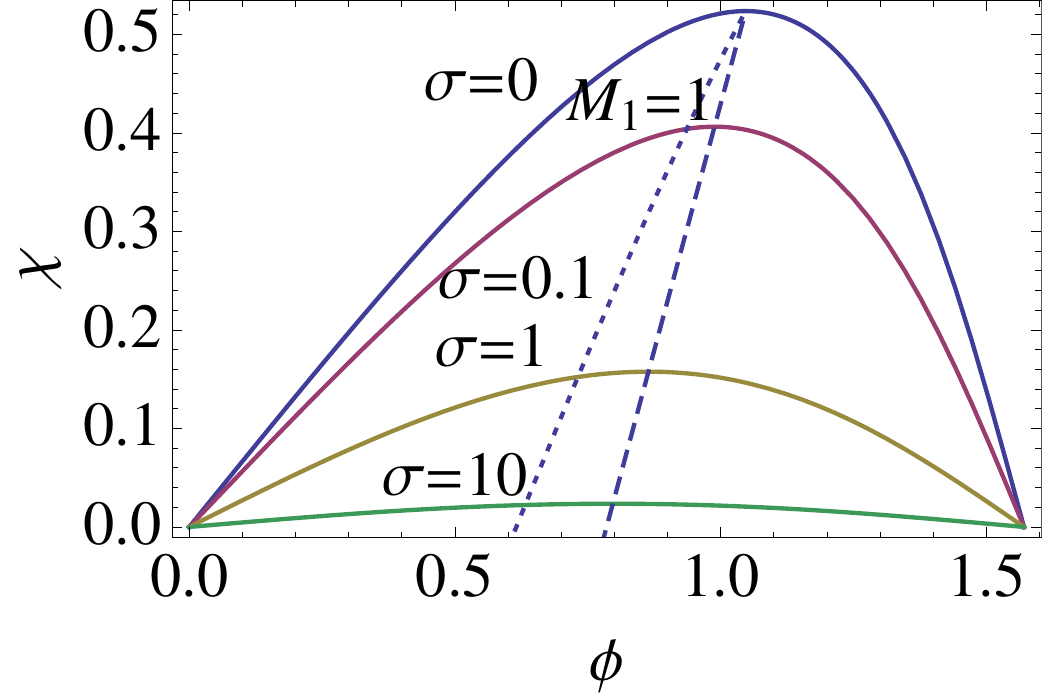}
\caption{Dependence of the deflection angle $\chi$ on the angle of attack $\phi$ for different magnetization parameters $\sigma$ in the limit $u_0^2 \rightarrow \infty$, . Dotted line is $M_1=1$ line, Eq. (\ref{MM2}); dashed line is the line of maximum deflection $\tan \chi_{max} = 
  1/2( \tan \phi -\cot \phi) $. }
\label{chi-sigma}
\end{figure}

   Using the post-shock  fast velocity \citep{1957PhRv..108.1357H}
   \be
   c_f^2 = { B_1^2 +4/3 p_1 \over B_1^2 + 4 p_1 + n_1},
   \ee the 
    post-shock
   Mach number is 
   \ba &&
   M_1^2={v_1 ^2 \over c_f^2}  =\frac{3 (  {\cal Z} +1) \left(\left(  {\cal Z} ^2-1\right) \sin ^2 \phi +1\right)}{7   {\cal Z} -1} 
   \approx 
   \nn &&
   \left\{ \begin{array}{ll}
 \frac{1}{3} +   \frac{8 \cos ^2\phi }{3}+  \left(\frac{2}{3}-\frac{56 \cos ^2\phi }{3}\right)\sigma , & \sigma \ll 1
   \\
1+   \frac{\cos ^2\phi -{2}/{3}}{\sigma }, & \sigma \gg 1
   \end{array}
   \right.
   \ea
   The flow becomes supersonic for
   \be
   \cos 2 \phi <  \frac{3  {\cal Z}^2+6  {\cal Z}-5}{3 ( {\cal Z}+1)^2}
   \label{MM2} \ee
   (see Fig. \ref{chi-sigma}).

   In particular, for strong fluid shocks,  $\sigma =0$, in the limit of high compression ratio, $\eta \gg 1$,
\ba && 
\eta^2 
= 8 u_0^2 \sin ^2 \phi 
\nn &&
u_1^2 = { 9 \over 8 \sin ^2 \phi } -1
\nn &&
\tan \chi = {\sin 2 \phi \over 2+ \cos 2 \phi }
\label{u23}
\ea
Maximum deflection  is reached at  $\phi = \pi/3 $ and  is equal to $\chi = \pi/6$ \citep{1980PhFl...23.1083K}.
In the large compression  limit the post-shock Mach number is
\be
M_1 ^2 ={3-(8/3) \sin ^2 \phi}
\ee
The flow becomes sonic at the maximum turning angle $\phi = \pi/3 $.

\end{document}